\documentstyle[aaspptwo,tighten]{article}

\begin{document}
\twocolumn
\title{Super-Giant Glitches and Quark Stars: Sources 
 of Gamma Ray Bursts?}

\author{Feng Ma and Bingrong Xie}
\affil{McDonald Observatory and Astronomy Department, University of Texas, Austin, TX 78712; 
feng, xiebr@astro.as.utexas.edu}

\begin{abstract}
When a spinning-down neutron star undergoes a phase transition that produces 
quark matter in its core, a Super-Giant Glitch of the order ${\Delta}
\Omega/\Omega\sim 0.3$ occurs on time scales from 0.05 seconds to a few
minutes. The energy released is about $10^{52}$ 
ergs  and can account for Gamma Ray Bursts at 
cosmological distances. The estimated burst frequency, 
$10^{-6}$ per year per galaxy, is in very good agreement
with observations. We also discuss the possibility of distinguishing  these events 
from neutron star mergers by observing the different temporal behavior of gravitational 
waves. 
\end{abstract}


\section{Introduction}

Although numerous explanations of Gamma Ray Bursts (GRBs) have been proposed, 
the exact nature of the GRB source, e.g., a neutron star binary merger (\cite{pac86}), 
a halo neutron star quake (Blaes, Blandford, \& Madau 1990), 
or a ``failed'' supernova (\cite{woosley93}), 
remains hidden behind a relativistically expanding
 fireball. Most observational consequences result from
radiation processes (see \cite{mes93} and 
\cite{th94} for ``generic'' models for GRBs) and are 
independent of the birth details. Hence, when 
considering the possible sources of GRBs, the essential parameters 
are the time scale, the initial
energy, the volume of the source, and most importantly, the birth rate of GRBs, 
which is about $10^{-6}$ per year per 
galaxy as indicated by observations (\cite{piran91}).

Here we outline how a neutron star might change into a hybrid star, which has a 
quark core and a neutron star crust (e.g., Rosenhauer et al. 1992). 
That is, a spinning-down neutron star increases in  central density
towards the critical
density ($\rho_{\rm cr} \sim 3\rho_0$, where 
$\rho_0 \simeq 2.8{\times}10^{14}$ g cm$^{-3}$ is the
nuclear density) for phase transition from neutron matter to quark 
matter (see \cite{Rosen} and references therein). 
Various Equations of State (EOSs) predict different central densities for 
a $1.4M_{\odot}$ neutron stars, and some neutron stars may have central
densities very close to $\rho_{\rm cr}$. They can evolve from the initial
situation with central density below the critical density to $\rho_{\rm cr}$ 
during the spin-down process. A phase transition occurring inside the
star causes it to collapse, thus releasing gravitational energy in the 
form of a GRB. A sudden spin-up, much more dramatic than any pulsar glitches observed, 
then takes place, which we call  ``Super-Giant Glitch (SGG)''. 

\section{Phase Transition and Super-Giant Glitch}

Baym and Chin (1976) studied the structure of a hybrid star, but they
considered  quark matter made of $u$ and $d$ quarks, which is 
stable only at densities higher than $10\rho_0$ and is unlikely to be 
reached in the center of neutron stars. Later, people found that
the strange quark matter  (made of approximately equal numbers of 
$u, d$ and $s$ quarks) has significantly lower energy than $u, d$ quark 
matter at the same pressure 
(\cite{farhi84}; \cite{Wit}). Witten even considered the possibility
that strange quark matter is more stable than $^{56}$Fe 
and is the absolute ground state of nature. 

With the conjecture that strange quark matter is stable at zero pressure,
some authors have studied the structure of strange stars (Alcock, Farhi \& Olinto 
1986a) and 
have even proposed a novel model which states that the GRB of 5 March 1979 was formed 
when a small lump of strange matter struck a rotating strange
star (\cite{Alcock86b}).  
It was also proposed (\cite{olesen91}) that a neutron 
star may burn into a strange star on time scales from 
0.05 seconds to a few minutes. The time scales depend mainly on 
the time scale for the weak interaction and a  diffusion coefficient 
(see \cite{olesen91} for details), and does not rely on assuming
whether strange quark matter is absolutely stable or not.

However, the existence of pulsar glitches (\cite{alpar87})  seems
 to be strong evidence against the existence of strange stars at all 
and, hence, the mechanism of GRBs from strange stars (\cite{kluz94}). 
Whether or not strange quark matter can be absolutely stable depends on
parameters like the bag constant in the quark model and is unclear today,  although
it is believed that strange quark matter is more stable than $u, d$ quark
matter and may be formed via deconfinement phase transition at a critical density 
of about $3\rho_0$.

Here we consider the case that strange quark matter is stable only at high
pressure with a critical density $\rho_{\rm cr}=3.0\rho_0$ for the phase transition. 
The mechanism of the conversion from a neutron star
to a hybrid star is that a small amount of quark matter is formed in the center
 once the central density of the neutron star
reaches the critical density, 
and since the EOS of the quark matter
is much softer than that of the neutron matter due to the asymptotic 
freedom property of quark-quark interactions, the newly formed quark 
matter cannot sustain the high pressure in the stellar center and will
be more highly compressed. As a result, the whole star  collapses  
until another stable configuration, a hybrid star, is reached. 

This point can also be seen from the mass-central 
density plot of hybrid stars (\cite{Rosen}),  where a discontinuous part in the 
curve indicates that a hybrid star cannot have a quark core of arbitrary size. 
A $1.4M_{\odot}$ hybrid star is stable only when it has a central density 
of about $10\rho_0$ and has more than half of its mass in the quark phase. 
Hence, once the central density of a neutron star reaches $\rho_{\rm cr}$, there
must be a sudden collapse.   Inside a hybrid star, the quark 
matter exists under high pressure, whereas in Witten's hypothesis
(\cite{Wit}) the strange matter is the absolute ground state 
and is stable at zero pressure. The mechanism of burning  
a neutron star to a strange 
star would be that a small lump of strange quark matter  ``eats up'' 
all the neutron matter.

There are numerous EOSs for the neutron matter. 
We use that of  Bethe \& Johnson (1974). It has very similar 
behavior to the phenomenological EOS of Sierk and Nix (1980) 
used in the detailed study of the structure of a static hybrid star by Rosenhauer et al (1992). 
A  $1.4M_{\odot}$  neutron star with Bethe \& Johnson's EOS has a radius of 
about 12 km (\cite{Shapiro}), while a hybrid star of the same mass has a radius 
of less than 10 km (\cite{Rosen}). 

The gravitational energy released in an SGG can be estimated by
\begin{equation}
 \displaystyle
    E \sim \frac{GM^2}{R}\left(\frac{{\Delta}R}{R}\right) \simeq 
10^{53} ~ \left(\frac{{\Delta}R}{R}\right) ~ {\rm ergs}, 
\end{equation} 
and is about $10^{52}$ ergs in our case. 
Most  of the dissipated energy is probably released in the form of neutrinos. 
If a small part of the total energy goes into $\gamma$ rays,  it will be large
enough to account for the GRBs at cosmological  distances
and to explain their isotropic distribution (\cite{palmer93}). 
Also, the phase transition and collapse can occur only 
once in a neutron star's life. This explains the lack of recurrence of GRBs. 

To estimate the spin-up rate of the star, we use the 
approximation of moment of inertia proposed by Ravenhall and 
Pethick (1994), 
\begin{equation}
 \displaystyle
  I\simeq 0.21MR^2\lambda(R)
    = 0.21\frac{MR^2}{1-2GM/Rc^2},\label{eq:iapp}
\end{equation}  
which is in fact a general relativistic correction to that of an 
incompressible fluid in the Newtonian limit $I=0.4MR^2$. Equation (2)  
is accurate to $10\%$ for EOSs without phase transitions, and 
to about $30\%$ for those EOSs predicting phase transitions.  
This is good enough in our order of magnitude estimate.

The moment of 
inertia of the neutron  star, according to equation (2), is 
$I_{\rm neutron} \simeq 1.3{\times}10^{45}$ g cm$^2$. 
The moment of inertia of the hybrid star is $I_{\rm hybrid} \simeq 1.0 \times 10^{45}$
 g cm$^2$. The change of angular velocity is then $\Delta\Omega/\Omega\sim 0.3$ and 
is $10^6$ times larger than the largest pulsar glitches observed.
The time scale for the SGG is similar to that for 
burning a neutron star into a strange star calculated by Olesen \& Madsen (1991),
and is from 0.05 seconds to a few minutes.
A more realistic description is complicated by the high temperature
during the SGG, while EOSs for finite temperature are very unclear. 
Also, the super-Eddington radiation may blow away the surface of the star.

\section{Burst Frequency}

In the spin-down process, the central density of a neutron star increases. 
The detailed study of Cook, Shapiro, \& Teukolsky (1994) shows that
the change of the central density (${\Delta}\rho_c$) is about 2.4\% for
a 1.4 $M_{\odot}$  neutron star spinning-down from an initial period of 4.3 
ms to a static state. 
We need to extrapolate from this to neutron stars with different spins.  
We note that the small
deviations of pressure and density from their equilibrium values have
 a rough relation of $dp/p \sim d\rho/\rho$ for a polytropic EOS, and that
the centrifugal force $\propto 1/P^2$, hence $d\rho/\rho \propto 1/P^2$.

We have very little knowledge about how fast a newborn neutron
star rotates, either theoretically or observationally. The only piece of information
available is that the Crab pulsar was born with a rotational period of 
about 20 ms (\cite{mt77}), so it will have a central density increase of 
about ${\Delta}\rho_c/\rho_c \simeq 0.001$ in its lifetime.

Assuming a phase transition critical density  $\rho_{\rm cr}$, 
only those neutron stars born with central densities 
$\rho_{\rm cr}(1-\Delta\rho_c/\rho_c) < \rho_c < \rho_{\rm cr}$
would have the chance to undergo the SGGs. For example, in the case of 
$\rho_{\rm cr} = 3.0\rho_0$, and assuming all neutron stars were born at the same initial
period ($P_i$) of 20 milliseconds, the density range is 
 $2.997\rho_0 < \rho_c < 3.0\rho_0$. Neutron stars with
lower central densities cannot reach the critical density in
their whole lives. Those with higher central densities should be
born as hybrid stars. 

Radio observations of binary pulsar systems together with statistics of 
neutron star mass distributions have given a strong constraint
on neutron star masses, which  lie in a  narrow
range from $1.0M_\odot$ to $1.6M_\odot$ (\cite{finn94}). 
For the EOS of Bethe \& Johnson (1974), the central densities of these neutron
stars range from a lower limit $\rho_{\rm l} \simeq 2.5\rho_0$ to an upper limit
$\rho_{\rm u} \simeq 4.3\rho_0$ (\cite{Shapiro}). It is apparent that we need EOSs predicting 
$\rho_{\rm l}<\rho_{\rm cr}<\rho_{\rm u}$ to make the sudden phase transition possible. 

We  assume the neutron star central densities are evenly distributed in this range. 
The birth rate of GRB events in units of per year per galaxy ($R_{\rm GRB}$)
 in our model will be the probability for a neutron 
star to undergo an SGG times the birth rate of neutron 
stars ($R_{\rm NS}$), 
\begin{equation}
 \displaystyle
    R_{\rm GRB}=\frac{\rho_{\rm cr}(\Delta{\rho_c}/\rho_c)}{\rho_{\rm u} - \rho_{\rm l}}~  R_{\rm NS} 
 \simeq 10^{-6} \left(\frac{P_i}{20{\rm\, ms}}\right)^{-2} \left(\frac{R_{\rm NS}}{10^{-3}}\right), 
\end{equation}  
where $R_{\rm NS}$ is an average over all types of galaxies, in units of per year per 
galaxy. $R_{\rm GRB}$ does not strongly  depend on the exact critical density 
since $\rho_{\rm cr}/(\rho_{\rm u}-\rho_{\rm l})$ most likely 
has an order of unity. With typical values of 
initial period and average neutron star birth rate the result from 
equation (3) is in very good agreement with observations. 

There are many factors affecting this estimate. 
First, some stars may be born as hybrid rather than 
neutron stars even with much lower central densities, 
because the high temperatures of  newborn neutron stars favor  the 
quark-hadron phase transition. Second, the stars may be born at  
rotational periods longer or shorter than 20 milliseconds. 
The uncertainties of the EOSs should also be kept in mind. There are numerous published
EOSs for high-density matter, 
while we only know that for a given density with a definite composition, there 
should be a correct one. A stiff mean field EOS (e.g., \cite{bp79}) predicts 
a 1.4 $M_\odot$ neutron star with $\rho_c \simeq 1.4\rho_0$, which does not
undergo an SGG even for the largest possible $\rho_c$ increase 
(30\% according to Cook et al. 1994);
while a soft EOS like that of Reid (e.g., \cite{bp79}) gives $\rho_c \simeq 10\rho_0$ 
and predicts that the star should be born as a hybrid star. 
If either of these EOSs is correct, there will be no SGGs at all. From equation (3), 
we can also give an upper limit for the SGG birth rate for the EOSs that favor 
the phase transition (like that of
Bethe \& Johnson).  With limiting values, $P_i \sim 0.5$ ms and  
$R_{\rm NS} \sim 0.02$ per year per galaxy, $R_{\rm GRB}$ can be as large as $10^{-2}$. 

\section{Discussions}

The predicted rate from the neutron star merger models is also close to
observations (\cite{piran91}), 
but these models have  difficulties like the inevitable disruption of 
the stars and  the rapid quenching
of the $\gamma$ ray emission due to the cooling and expansion of the 
ejected baryonic matter.
In our model, there is less possibility of  disruption.  
Hence, the problems of rapid quenching and 
contamination are minimized. On the other hand, the SGGs offer much
more energy than ordinary starquake models (\cite{bb}). In the 
latter models, the GRBs are interpreted as events within our own Galaxy 
and have apparent difficulties in explaining the observed isotropic 
distribution. The SGG model proposed here 
is rather natural, since most pulsars are spinning-down and increasing
their central densities. We do not have to assume stable strange matter
or other exotic and rare events. The phase transitions inside neutron 
stars are not solely quark-hadron
phase transitions. Other phase transitions like  
pion condensation  may also result in SGGs and account for GRBs. 

If an SGG happens late in 
the spin-down history  of a neutron star, it is  a nearly spherical 
collapse and produces no  gravitational radiation. 
Most of the observed millisecond pulsars are believed to be old and have been spun up by
accretion. They have weak magnetic fields and low spin-down rates which do not 
favor SGGs. However, it is possible that some neutron stars may be born at high
initial spins (Michel 1987; Lai \& Shapiro 1995). If an SGG happens in 
a neutron star with rotational period about 1 millisecond, it may
produce detectable quadrupole gravitational radiation. The collapse with rotation 
is similar to  the lowest quadrupole mode of vibration of 
a rotating neutron star for which the power of gravitational radiation has been
estimated for a vibration amplitude of $0.1R$ (which is
approximately equal to the stellar surface displacement in our SGG model) and 
a rotational period of 1 millisecond (\cite{wheeler66}; \cite{mis73}). 
The power is about  
 $10^{50}$ ergs sec$^{-1}$ and is comparable to that in a 
neutron star merger (\cite{koc93}; \cite{cen94}; \cite{lip95}), 
but is apparently  associated with different waveforms. So it is possible 
to discriminate the SGGs from neutron star mergers  with gravitational wave detectors
like the Laser Interferometer Gravitational Wave Observatory (LIGO; see Abramovici 
et al. 1992).


In conclusion, we propose that the sudden transformations from 
neutron stars to
hybrid stars may account for the Gamma Ray Bursts at cosmological
distances. We also give explanations to 
the properties of GRBs: the duration of bursts, burst frequency, the
lack of recurrence and the isotropic distribution. If this model is
eventually confirmed in further observations, in addition to improving 
our understanding of GRBs, the existence of quark 
matter can be proved. Otherwise, the EOSs and the 
parameters related to the phase transition in this $Letter$ will 
be challenged, although they have been widely studied in literature. 
In this way, we may be able to give constraints on some EOSs and/or 
parameters like the bag constant in the quark model, which is essential 
in determining  the critical density for the quark-hadron phase 
transition.

\acknowledgements 

We thank Robert Duncan for helpful discussions; Byron Mattingly and 
Erik Gregersen for help 
with the manuscript. We are especially grateful to the referee, Patrick Mock, 
whose numerous valuable suggestions made the current form of this paper possible. 

\end{document}